\documentclass{aip-cp}

\usepackage[numbers]{natbib}
\usepackage{rotating}
\usepackage{graphicx}
\usepackage{caption}

\begin{document}

\title{The Medium Size Telescopes of the Cherenkov Telescope Array}

\author[aff1]{G. P\"uhlhofer\corref{cor1}}
\author[aff2]{the CTA Consortium}

\affil[aff1]{Institut f\"ur Astronomie und Astrophysik, Eberhard Karls Universit\"at T\"ubingen, Sand 1, 72076 T\"ubingen, Germany}
\affil[aff2]{see http://www.cta-observatory.com for full author and affiliation list}
\corresp[cor1]{Corresponding author: Gerd.Puehlhofer@astro.uni-tuebingen.de}

\maketitle

\begin{abstract}
The Cherenkov Telescope Array (CTA) is the planned next-generation instrument for ground-based gamma-ray astronomy, covering a photon energy range of $\sim$20\,GeV to above 100\,TeV. CTA will consist of the order of 100 telescopes of three sizes, installed at two sites in the Northern and Southern Hemisphere. This contribution deals with the 12 meter Medium Size Telescopes (MST) having a single mirror (modified Davies-Cotton, DC) design. In the baseline design of the CTA arrays, 25 MSTs in the South and 15 MSTs in the North provide the necessary sensitivity for CTA in the core energy range of 100\,GeV to 10\,TeV. DC-MSTs will be equipped with photomultiplier (PMT)-based cameras. Two options are available for these focal plane instruments, that will be provided by the FlashCam and the NectarCAM sub-consortia. In this contribution, a short introduction to the projects and their status is given.
\end{abstract}

\section{INTRODUCTION}
The components for the CTA \cite{bib:Design2011} baseline MSTs will be constructed through three sub-projects, the MST structure, the FlashCam, and the NectarCAM projects. Currently, 8 countries with 27 institutes and of the order of 125 persons are involved in these three projects. CTA baseline is to deploy 25 MSTs at the Southern site (foreseen to be hosted by ESO close to the Paranal site in Chile) and 15 MSTs at the Northern site, hosted by IAC at La Palma (Canary Islands, Spain). In the initial stage of the CTA project for which funding is currently being secured, 15 MSTs need to be delivered to Chile and 5 MSTs to La Palma. 

In the following, the status and progress of the three sub-projects is briefly sketched, giving also references to more detailed descriptions of the sub-components presented at this or earlier conferences.

\section{The Medium Size Telescope structure}

\begin{figure}[htb]
  \centerline{\includegraphics[width=0.8\textwidth]{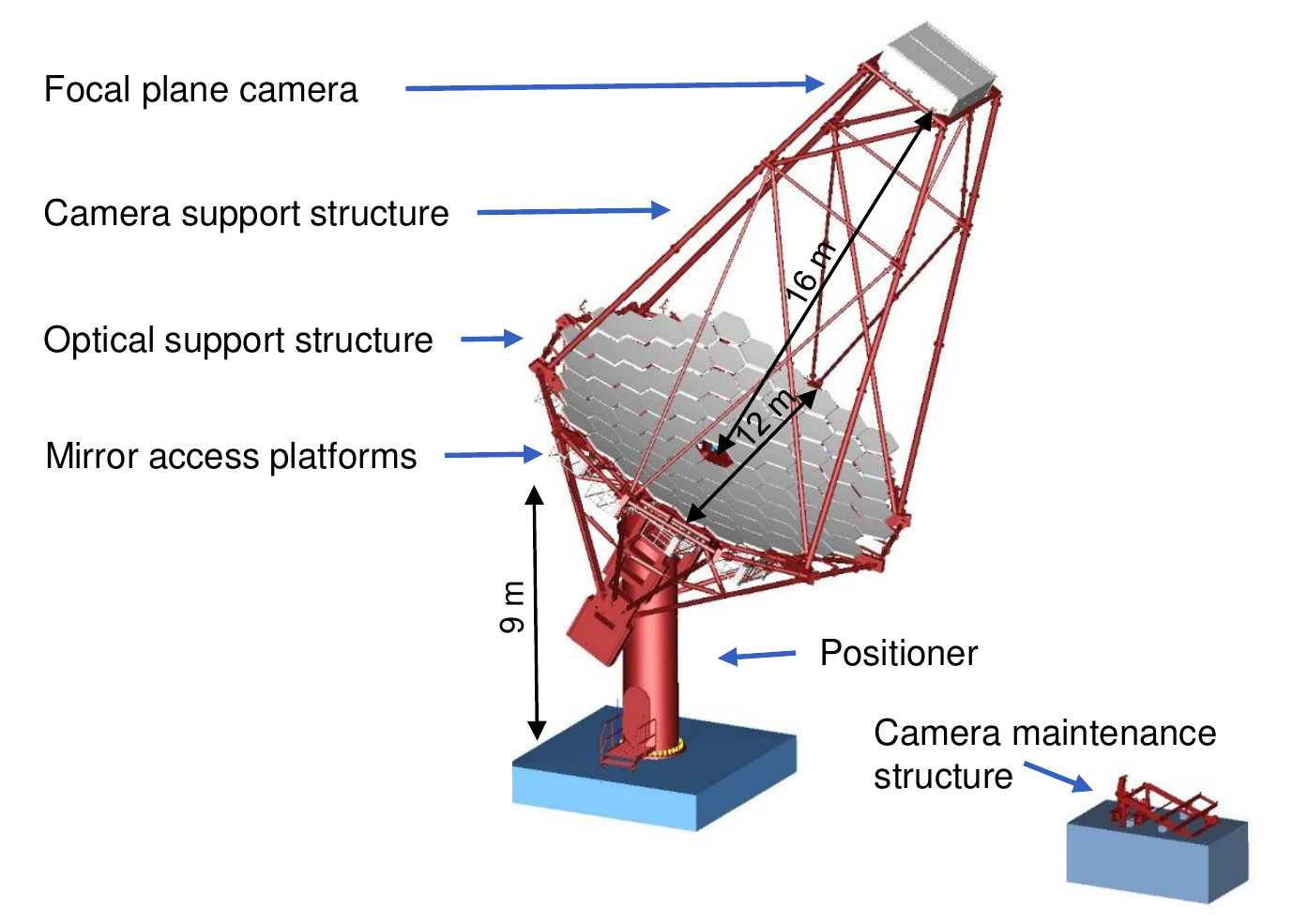}}
  \caption{Design of the MST structure.}
  \label{Fig:MSTstructure}
\end{figure}

\begin{figure}[htb]
  \centerline{\includegraphics[width=0.8\textwidth]{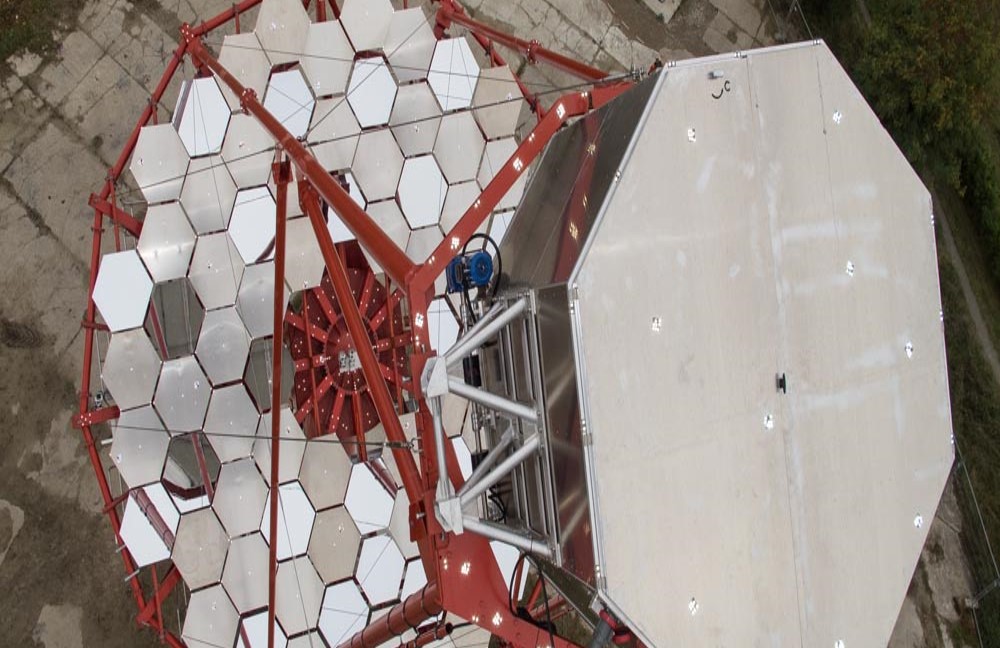}}
  \caption{The MST structure prototype in Berlin-Adlershof, viewed in direction of the incoming light. On the picture, the focal plane is equipped with a dummy camera, and mirrors from different prototyping stages as well as dummy mirrors are mounted at the dish.}
  \label{Fig:MSTprototype}
\end{figure}

The MST structures \cite{bib:MSTStructureICRC2015} (see Fig.\,\ref{Fig:MSTstructure}) are made of steel to ensure sufficient stiffness of the optical support structure under all elevation angles. The mirror is tesselated and consists of 86 hexagonally-shaped mirrors, with 1.2\,m flat-to-flat diameter. The optics of the MST is based on a modified Davies-Cotton design. The dish has a radius of curvature of 19.2\,m, while the focal plane of the telescope is located at a distance of 16\,m from the dish center. The MST optics improves the isochronocity without strong impact on the ideal DC point spread function. The radius of curvature of the spherical mirror tiles themselves is 32.14\,m, i.e.\ approximately twice the focal length, optimized to give the best point spread function during regular wobble-mode observations. The mirrors are aligned to image rays along the optical axis into the focal point. Individual mirrors are mounted to the dish while being attached to their support triangles. The dish features mirror access platforms at the back for easy access. The positioner of the telescope is designed as a central tower, which has three floors that can be accessed by maintenance personnel and during construction. Besides the drive system, the tower hosts all electrical cabinets relevant for the MST structure. The total weight of the telescope, including the focal plane camera, is of the order of 86 tons. 

A prototype of the MST telescope structure has been built in Berlin-Adlershof (see Fig.\,\ref{Fig:MSTprototype}) and has been used extensively over the last years to gain experience with the MST design and develop improvements where necessary. Results of the experience with the drive system as well as with the overall design including its optimization are reported in \cite{bib:MSTStructureGamma2016} (these proceedings). Calibration techniques that have been developed and tested at the structure prototype are discussed in \cite{bib:MSTCalibrationGamma2016} (these proceedings), including pointing calibration and procedures for mirror alignment.

After mounting, mirror tiles (that will be delivered by different providers to ensure sufficient capacities during mass production) are aligned by means of motorized actuators which are supporting all mirror tiles individually. For initial alignment, the Bokeh method \cite{bib:Ahnen2016} is used. A fine-alignment of all tiles is performed in a second step, using stars as reference objects while tracking the telescope. The dish is designed such that the point spread function is within requirements under all operational elevation angles, without the need for realigning mirrors. Nevertheless, the actuator system is designed such that it could adapt the individual mirror alignment under different elevation angles to compensate slight elastic deformations if ultimately necessary. The alignment system can also be used to realign mirrors on timescales of months to years to compensate mechanical relaxation if occuring. A choice for an actuator system specifically optimized for MST needs is described in \cite{bib:MSTActuatorsGamma2016} (these proceedings). An actuator system currently under central CTA development for all CTA telescope types, combining experience of several CTA developments, may in the end be the optimal choice.

The telescope dish also hosts auxiliary devices like camera calibration devices that are designed and controlled by the focal plane instrumentation teams, and (optical) CCD cameras that are used for mirror alignment and for pointing calibration of the telescope. The telescope drive system is capable of tracking celestial objects stably (using active vibration damping) with an accuracy of $<$$2''$ RMS. The overall tracking precision (including elastic telescope deformations) is better than $0.1^{\circ}$, which is sufficient for online tracking due to the short event integration times. The required offline accuracy (of better than $7''$ in the case of MSTs), matching the position of the field of view (FoV) to the tracked stellar field at any time, can however not be provided by the PMT-based Cherenkov telescope cameras. Guiding systems are therefore necessary to monitor the position of the dish and of the focal plane camera with respect to the sky (while the position of the optical axis defined by the mirror tiles with respect to the dish can only be calibrated offline and therefore relies on the mechanical stability of the mirror support system). Problems of relative calibration of two CCD cameras monitoring the sky and the focal plane camera separately can be overcome by means of a specifically designed single-CCD system which is described in more detail in \cite{bib:SingleCCDGamma2016} (these proceedings).

\section{Focal plane instrumentation}

On the ground, next to the telescope, more auxiliary devices like the focal plane camera maintenance structure (cf.\ Fig.\,\ref{Fig:MSTstructure}) and liquid cooling and dry air systems for the focal plane cameras are located. The focal plane cameras themselves employ high quantum-efficiency PMTs to collect the incoming Cherenkov light of the air-shower events with integration times on nanosecond scale. The readout and trigger electronics are located in the focal plane camera housings as well. Behind an acrylic entrance window, the light is concentrated on the PMT cathodes by means of funnel plates, made of (hollow) Winston cones. Cameras have a FoV of above $7.5^{\circ}$ diameter, $>$$1750$ pixels are arranged in a hexagonal matrix of $50\,\mathrm{mm}$ spacing, corresponding to $0.18^{\circ}$ pixel size in the MST focal plane.

The cameras are mechanically supported in the primary focus of the telescope by a steel-based camera-support structure with an aluminium frame around the cameras, and are interfaced only by power, liquid cooling and dry air pipes, and by optical fiber cables for control and bulk data transfer. Interfaces are designed such that both currently available focal plane cameras fit into the telescope. The cameras are described in more detail in the following sections.

\section{FlashCam}

\begin{figure}[ht]
  \centerline{\includegraphics[width=0.8\textwidth]{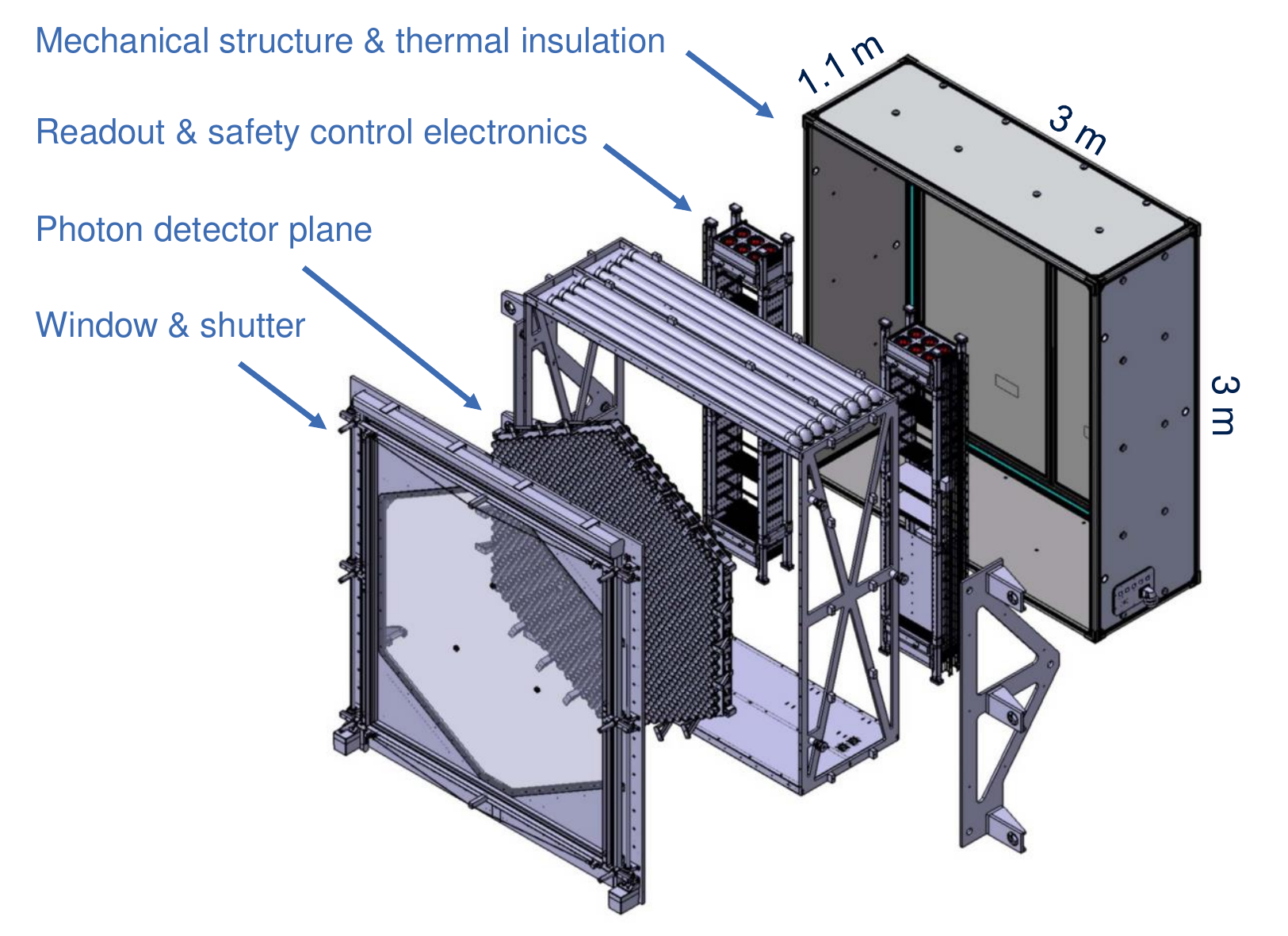}}
  \caption{Exploded view of the FlashCam structural design. The readout electronics arranged in crates as well as the photon detector plane units can be seen.}
  \label{Fig:FlashCamDrawing}
\end{figure}

\begin{figure}[ht]
  \centerline{\includegraphics[width=0.36\textwidth]{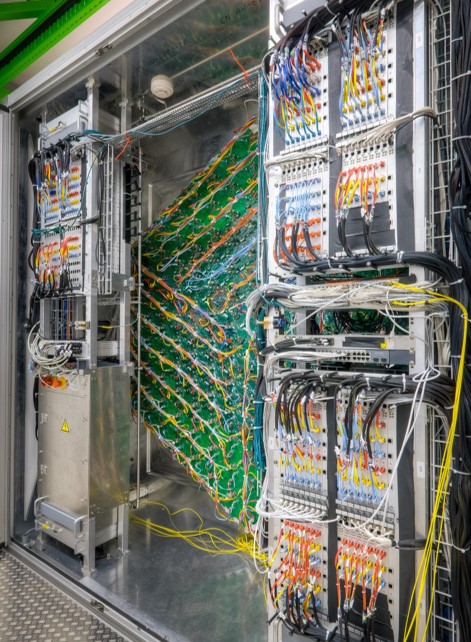}
  \hspace{7mm} \includegraphics[width=0.5\textwidth]{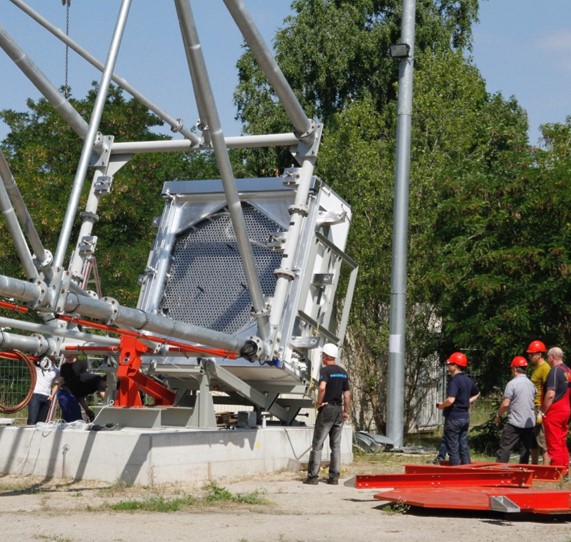}}
  \caption{Left panel: The FlashCam prototype viewed from behind with open doors. 
  Right panel: The FlashCam prototype mounted at the MST structure prototype for interface tests.}
  \label{Fig:FlashCamPrototype}
\end{figure}

\begin{figure}[h]
\minipage{0.82\textwidth}
\centering
  \includegraphics[width=0.8\linewidth]{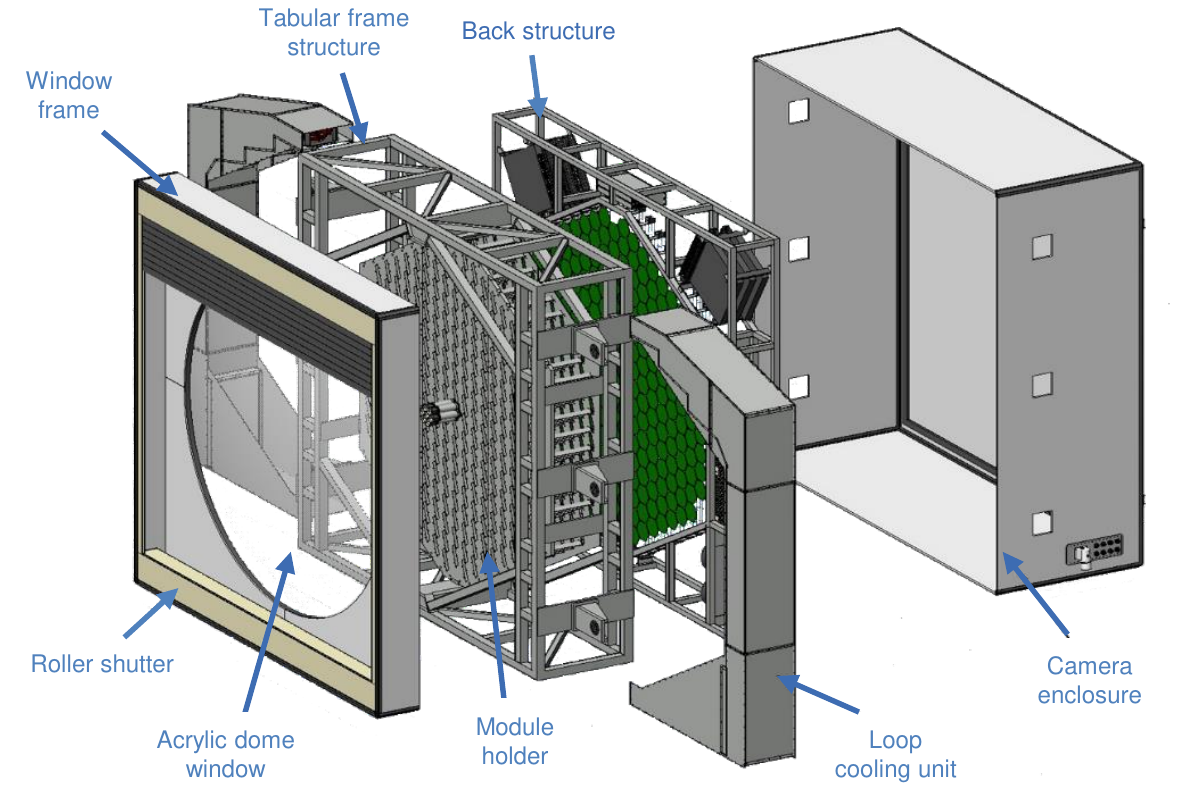}
~\\
~\\
~\\
  \includegraphics[width=0.8\linewidth]{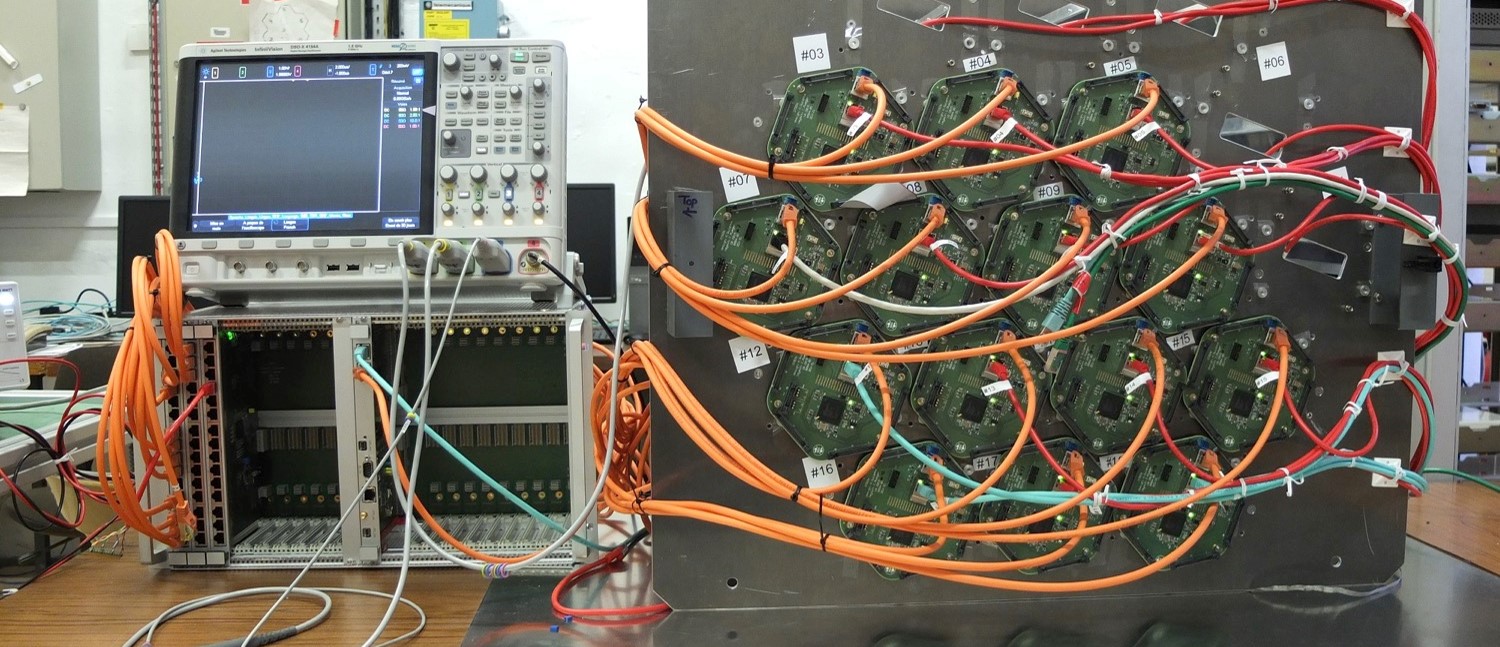}
\endminipage\hfill
\caption{Top panel: Exploded view of the NectarCam structural design. Bottom panel: 19-module NectarCam prototype setup to evaluate the performance of the different trigger logics.}
\label{Fig:NectarCam}
\end{figure}

The FlashCam cameras feature as one of their core elements a fully digital trigger and readout system that has been designed, prototyped and successfully tested by the FlashCam team over the past years. Cherenkov telescope cameras need to create an event trigger from the incoming light distribution by themselves. Short timescales are necessary to minimize dead-time. In FlashCam, PMT signals are (after pre-amplification and shaping) digitized using commercial low-power 250\,MHz pipeline-ADCs, and trigger creation (as well as further pre-processing of the signals) is performed in the front-end electronics by commercial low-cost FPGA, solely based on these digitized signals. The design (including a preamplifier that behaves deterministically non-linearly at high incoming charges) permits to digitize the full dynamic range of a pixel ($\sim$0.2 -- 3000 photo-electrons [p.e.]) in one channel, saving costs as well as bandwidth of the subsequent data transfer. 

Figure \ref{Fig:FlashCamDrawing} shows a schematic view of a FlashCam MST camera, hosting 1758 pixels. Photon detector modules (each with 12 PMTs including preamplification circuits, HV supply, and control/monitoring) are physically detached from the readout front-end electronics. The readout electronics boards (serving 24 channels each) as well as trigger and master distribution boards are organized in crates and racks towards the back of the camera. The racks can be accessed for installation and maintenance from behind the camera after opening the rear doors. Photon detector plane modules are installed (and exchanged in case of maintenance repair) from the inside of the camera, without the need to remove the optical front system. Cooling inside the camera is performed by means of dry air flow, the dissipated $\sim$$4.5\,\mathrm{kW}$ is brought to ground by the liquid cooling system mentioned before. The camera housing is slightly overpressured to avoid dust entering through possible gaps in the enclosure.

Figure \ref{Fig:FlashCamPrototype}, left panel shows a rear view of the FlashCam camera prototype, that has been operational since October 2015 in a dark room for extensive testing, after an interface campaign at the MST structure prototype (right panel of Fig.\,\ref{Fig:FlashCamPrototype}). Bulk data transfer from the front-end to a dedicated server computer over 1\,km 4$\times$10\,G fibers has been successfully demonstrated with a dead-time free event rate of $>$$20\,\mathrm{kHz}$. Using a highly efficient ethernet protocol developed for this purpose, event traces of $\sim$2000 channels have been transmitted with full resolution. 

The prototype has been used to repeat the verification of performance parameters such as amplitude and time resolution under realistic night-sky background conditions, with $\sim$half camera-scale number of pixels. Specifically, two different choices ($\sim$350 each) of PMTs, developed by Hamamatsu for CTA needs, have been implemented and extensively tested regarding afterpulsing, demonstrating that all PMTs are within requirements ($<2\times10^{-4}$ above 4\,p.e.). Furthermore, the prototype undergoes mechanical stress tests as well as performance tests under different temperature conditions.

The FlashCam design as well as prototyping results and status updates have been described in a series of conference papers, which are referred to for further details \cite{bib:Gamma2012,bib:ICRC2013,bib:SPIE2014,bib:ICRC2015}. Aspects for electronics test procedures towards mass production have been shown in \cite{bib:FlashCamTestsGamma2016} (these proceedings).

\section{NectarCAM}

The NectarCAM camera for MSTs is designed around the Nectar analog pipeline readout chip. Incoming PMT signals are continuously sampled at sampling speeds of 0.5 -- 2\,GHz (default is 1\,GHz). After an event trigger produced inside the camera trigger logic (which is derived from a separate electronics path combining pixel charges), sampling is stopped and signals are digitized and sent by the front-end electronics to a camera server computer. In order to provide the full dynamic range (0.5 -- 2000\,p.e.), each pixel signal is sampled with two channels, a low-gain and a high-gain channel. Efficient triggering and readout components ensure a low dead-time of $<$3\% at $4.5\,\mathrm{kHz}$ event rate for a full MST camera.

Figure \ref{Fig:NectarCam} shows a schematic view of a NectarCAM camera, hosting 1855 pixels. The PMTs and readout electronics are arranged in modules of 7 pixels each, that are installed (and exchanged in case of maintenance repair) from the front side of the camera. Individual modules have been extensively tested and verified for their performance. The total heat dissipation in the camera is around $7.7\,\mathrm{kW}$. Inside the camera, dry air flow is used to cope with the heat, together with the exterior liquid cooling system for heat exchange to the outside as discussed above. The cooling system has been qualified with a demonstrator system.

At the backside of the PMT/electronics modules, modular trigger electronics boards are connected. Since the trigger path is detached from the sampling path, both an analog trigger (creating analog trigger sums) and a digital trigger scheme (adding pixel trigger information after simple pixel digitization) are possible. Both trigger schemes are still possible within the current NectarCAM design, prototypes for both systems are available and have been extensively tested in a 19 module camera test setup (see Fig.\,\ref{Fig:NectarCam}).

Bulk data transfer from the NectarCAM front-end electronics to a camera server is demanding due to the high pixel data rate, if full event waveforms from all pixels are transmitted for each event. Initial considerations to possibly only transmit a fraction of the FoV per event (where the bulk of the event information is identified on an event basis) have been superseded, since in the meantime full event transmission between front-end and server has been demonstrated on dedicated test setups using simulated data.

More details of the NectarCAM design and the current prototyping status have been presented in \cite{bib:NectarCAMGamma2016} (these proceedings).

\section{Outlook}

The MST structure, FlashCam and NectarCAM teams are currently preparing towards pre-production telescope structures and cameras, to be deployed on the final CTA sites when they become available. Current plans foresee two telescopes with FlashCam cameras at the Southern site to be ready in mid 2018, and one telescope equipped with a NectarCAM camera at the Northern site on similar timescales.

\section{ACKNOWLEDGMENTS}
The author thanks S. Schlenstedt and M. Garczarczyk, J.-F. Glicenstein and M. Fesquet, and G. Hermann from the MST structure, NectarCAM, and FlashCam projects for providing material for the presentation and for reviewing the manuscript.
We gratefully acknowledge support from the agencies and organizations under Funding Agencies at www.cta-observatory.org.

\nocite{*}
\bibliographystyle{aipnum-cp}%

\end{document}